\documentstyle[11pt,epsfig]{article}
\newcommand{\be}{\begin{equation}}
\newcommand{\ee}{\end{equation}}
\newcommand{\abz}{\hspace*{.5in}}

\refstepcounter{equation}

\begin{document}
\title{Exact Einstein--scalar field solutions for formation of black holes
in a cosmological setting}
\author{Oleg A. Fonarev
\\ {\em Racah Institute of Physics, The Hebrew University}
\\ {\em Jerusalem 91904, Israel}}
\date{}

\maketitle
\begin{abstract}
We consider self--interacting scalar fields coupled to gravity.
Two classes of exact solutions to Einstein's equations are obtained: the first
class corresponds to the minimal coupling, the second one to the conformal
coupling. One of the solutions is shown to describe a formation of a black
hole in a cosmological setting. Some properties of this solution are described.
There are two kinds of event horizons: a black hole horizon and cosmological
horizons. The cosmological horizons are not smooth. There is a mild curvature
singularity, which affects extended bodies but allows geodesics to be extended.
It is also shown that there is a critical value for a parameter on which the
solution
depends. Above the critical point, the black hole singularity is hidden within
 a global black hole event horizon.
Below the critical point, the singularity appears to be naked. The relevance
 to cosmic censorship is discussed.
\end{abstract}
\vspace{2\baselineskip}
PACS numbers: 04.20.Cv, 04.20.Jb, 97.60.Lf

\thispagestyle{empty}
\setcounter{page}{0}
\newpage
\setcounter{page}{1}

\section{Introduction}
\setcounter{equation}{0}
\abz
Two remarkable predictions of General Relativity, the expanding universe
and black holes, have always been of main interest in the theory of
gravitation.
In fact, the expansion of the universe and the gravitational collapse of
matter are very similar phenomena from the mathematical point of view
\cite{kn:hawking}.
Both of them yield the existence of regions in the space--time where
matter has transplanckian energies.
It is clear that classical General Relativity can not be extrapolated there.
What theory describes the early stage of the evolution
of our universe and the final stage of the gravitational collapse?
Will the notions of space and time survive in a future theory of Quantum
Gravity at all?
Or more specifically: what is left after a black hole has evaporated
and what was before the Big Bang?
These questions are among the most fundamental issues  in
contemporary theoretical physics. \\ \abz
Yet, there are many unresolved problems in the classical General Relativity.
One of such problems is the dearth of exact solutions describing collapsing
matter and formation of black holes.
While many cosmological models are available in the market, very few
{\it dynamical} solutions
are known in (4D) black hole physics - the most popular are the Tolman solution
\cite{kn:tolman} for
the collapse of a dust and the Vaidya solution \cite{kn:vaidya} for the
collapse of radiation.
Some recent numerical studies of different models of the gravitational
collapse \cite{kn:choptuik,kn:abraham} indicate that
black hole formation may be accompanied with other interesting phenomena,
such as echoing and critical behaviour.
In view of this, it would be extremely important to have exact analytic
solutions of Einstein's equations which could be regarded as describing
collapsing matter.
This paper aims at obtaining such solutions. \\ \abz
We shall study two models of self--interacting scalar fields.
In the first model the scalar field is minimally coupled to gravity and the
scalar potential has a Liouville form, $V \propto \exp (-k \varphi)$.
This model has been studied by numerous authors with a view to cosmological
applications (see, for instance, Refs.
\cite{kn:halliwell, kn:barrow, kn:feinstein}).
The Liouville potential arises as an effective potential in some supergravity
theories or in Kaluza--Klein theories after dimensional reduction to an
effective 4--dimensional theory \cite{kn:halliwell}.
It also arises in higher--order gravity theories after a transformation
to the Einstein frame \cite{kn:barrow}. \\ \abz
In the second model the scalar field is conformally coupled to gravity.
This model is less studied, though it was proved by Bekenstein
\cite{kn:bekenstein1} that there is a mapping between the space of
Einstein--conformal scalar solutions and the space of Einstein--ordinary
scalar solutions.
Bekenstein found the mapping for massless scalar
fields but the result can easily be extended to self--interacting fields
\cite{kn:abreu}.
It turns out that the counterpart of the ordinary--scalar Liouville potential
 is the following conformal--scalar potential:
$W \propto (1- \zeta \phi)^{2+k/(2 \zeta)} (1 + \zeta \phi)^{2-k/(2 \zeta)}$,
where $\zeta = (4 \pi G/3)^{1/2}$.
The most interesting case appears to be $k = \pm 2 \zeta$. In this case we find
a
solution of the Einstein--conformal scalar equations which is shown to
describe the gravitational collapse of the scalar field in a cosmological
setting. \\ \abz
This paper is organized as follows. In Sec.\ ~\ref{sec-theorems} we prove
two theorems that allow one to generate new exact solutions of the Einstein--
scalar field equations from static vacuum solutions of Einstein's equations.
As an application of the theorems we obtain spherically symmetric dynamical
solutions in the two models mentioned above.
The cosmological black hole solution is studied in Sec.\ ~\ref{sec-bh}.
Section ~\ref{sec-conclusion} contains some concluding remarks.
Sign conventions for the metric and the Riemann tensor follow those of the
book \cite{kn:landau}.

\section{Generating new solutions of the Einstein--scalar equations}
\setcounter{equation}{0}
\label{sec-theorems}
\subsection{Minimal coupling}
\label{sec-theorem1}
\abz Let us consider the Einstein--scalar system described by the action
\be
{S_{m}}[g_{\alpha\beta},\varphi] = \int d^{4}\,x\: (-g)^{1/2} \left[
-(16 \pi G)^{-1} R + \frac{1}{2} \varphi_{, \alpha} \varphi_{, \beta}
g^{\alpha\beta} - {V}(\varphi) \right] \; , \label{eq:minaction}
\ee
with an exponential potential,
\be
{V}(\varphi) = V_{o} \exp (-k \varphi) \; . \label{eq:exppot}
\ee
Variations of the action with respect to $g^{\mu\nu}$ and $\varphi$ yield
the following equations:
\be
R_{\mu\nu} = 8 \pi G \left( \varphi_{, \mu} \varphi_{, \nu} - g_{\mu\nu}
V_{o} \exp (-k \varphi) \right) \; , \label{eq:einmin}
\ee
\be
\Box \varphi = k V_{o} \exp (-k \varphi) \; . \label{eq:box}
\ee
Our first theorem is:
\newtheorem{guess}{Theorem}
\begin{guess}
If $ds_{o}^{2} = e^{2 u} dt^{2} - e^{-2 u} h_{ij}
dx^{i} dx^{j}$ - is a static vacuum solution of Einstein's equations, then
\be
ds^{2} = \exp (2 \delta u + 8 \alpha^{2} a T) dT^{2} -
\exp (-2 \delta u +2 a T) h_{ij} dx^{i} dx^{j} \; , \label{eq:minmetric}
\ee
\be
\varphi = (4 \pi G)^{-1/2} \left( (4 \alpha^{2} + 1)^{-1/2} u + 2 \alpha
a T \right) \label{eq:minphi}
\ee
form a solution of equations (\ref{eq:einmin}) and (\ref{eq:box}), where
the parameters $\alpha, \delta$ and $a$ are related to the parameters of
the potential as follows:
\be
\begin{array}{rll}
k & = & 8 \alpha (\pi G)^{1/2} \; , \\
\delta & = & 2 \alpha (4 \alpha^{2} + 1)^{-1/2}\; , \\
V_{o} & = & (8 \pi G)^{-1} a^{2} (3 - 4 \alpha^{2}) \; . \end{array}
\label{eq:parameters}
\ee
\end{guess}
{\it Proof.} For $a=0$ (no potential term) the corresponding theorem was
proved in \cite{kn:buchdall,kn:janis}.
Consider thus the static solution of equations (\ref{eq:einmin}) and
(\ref{eq:box}) with $V_{o}=0$,
\be
d\bar{s}^{2} = \exp (2 \delta u) dT^{2} -
\exp (-2 \delta u) h_{ij} dx^{i} dx^{j} \; , \label{eq:barmetric}
\ee
\be
\overline{\varphi} = (4 \pi G)^{-1/2} (4 \alpha^{2} + 1)^{-1/2} u \; .
 \label{eq:barphi}
\ee
Under the conformal transformation,
\be
g_{\mu\nu} = \exp (2 {\mu}(t)) \, \overline{g}_{\mu\nu} \; , \label{eq:conf}
\ee
the Ricci tensor transforms as follows (see, for instance, \cite{kn:yano})
\be
\begin{array}{l}
R_{oo} = \bar{R}_{oo} - 3 \ddot{\mu} \; , \\
R_{oi} = \bar{R}_{oi} + 2 \dot{\mu} \partial_{i} \log (g_{oo}) \; , \\
R_{ij} = \bar{R}_{ij} - ( \ddot{\mu} + 2 \dot{\mu}^{2} )
g_{ij} g^{oo} \; . \label{eq:ricconf}
\end{array}
\ee
Since (\ref{eq:barmetric}) and (\ref{eq:barphi}) satisfy the Einstein
equations,
\be
\bar{R}_{\mu\nu} = 8 \pi G \overline{\varphi}_{\mu} \overline{\varphi}_{\nu}
\; , \label{eq:einbar}
\ee
it is easy to check that the metric tensor (\ref{eq:conf}) and the scalar
field
\be
\varphi = \overline{\varphi} + \alpha (\pi G)^{-1/2} {\mu}(t)
\label{eq:confphi}
\ee
satisfy equation (\ref{eq:einmin}) if the function ${\mu}(t)$ obeys the
following
two equations:
\be
\ddot{\mu} = (1-4 \alpha^{2}) \dot{\mu}^{2} \; , \label{eq:mu1}
\ee
\be
(3 - 4 \alpha^{2}) \dot{\mu}^{2} = 8 \pi G V_{o} \exp (2 (1-4 \alpha^2) \mu)
\; . \label{eq:mu2}
\ee
When $\alpha^{2} \neq 3/4$, equation (\ref{eq:mu2}) implies equation
(\ref{eq:mu1}) (the Bianchi identity \cite{kn:hawking}).
If $\alpha^{2} = 3/4$, equation (\ref{eq:mu2}) is trivial. \\ \abz
There are two distinct cases:
\begin{enumerate}
\item
$ \alpha^{2} = 1/4 $. The solution of equations (\ref{eq:mu1}) and
(\ref{eq:mu2}) is ${\mu}(t) = a t + \mu_{o}$. Taking into account equations
(\ref{eq:conf}) and (\ref{eq:confphi}), one immediately gets the solution
(\ref{eq:minmetric})--(\ref{eq:parameters}).
\item
$ \alpha^{2} \neq 1/4 $. In this case the solution is
${\mu}(t) = (4 \alpha^{2} -1)^{-1} \log |(4 \alpha^{2} -1) a t| + \mu_{o}$.
Redefining the time coordinate,
$(4 \alpha^{2} -1) a t = \pm \exp ((4 \alpha^{2} -1) a T)$, we arrive at
equations (\ref{eq:minmetric})--(\ref{eq:parameters}).
\end{enumerate}
\abz To complete the proof we notice that equation (\ref{eq:box}) coincides,
in our case, with the (0,0)--component of Einstein's equations. Q.E.D. \\
\\
\abz
As an application of the theorem we shall consider spherically symmetric
solutions of equations (\ref{eq:einmin}) and (\ref{eq:box}).
The corresponding vacuum metric is the Schwarzschild metric:
\be
e^{2 u} = 1 - 2 m/r \; , \label{eq:uschw}
\ee
\be
h_{ij} dx^{i} dx^{j} = dr^{2} + (1-2 m/r) r^{2} ( d\theta^{2} + \sin^{2} \theta
d\varphi^{2} )\; . \label{eq:hschw}
\ee
Applying our theorem, we get, thus, the following solution of equations
(\ref{eq:einmin}) and (\ref{eq:box}):
\begin{eqnarray}
ds^{2} & = & \exp (8 \alpha^{2} a T) (1-2 m/r)^{\delta} dT^{2} \nonumber \\
 & & \mbox{} - \exp (2 a T) \left( \frac{dr^{2}}{(1-2 m/r)^{\delta}} +
(1-2 m/r)^{1-\delta} r^{2} ( d\theta^{2} + \sin^{2} \theta
d\varphi^{2} ) \right) \; , \label{eq:mindssol}
\end{eqnarray}
\be
\varphi = (4 \pi G)^{-1/2} \left( 2 \alpha a T +
(4 \alpha^{2} + 1)^{-1/2} \frac{1}{2} \ln (1-2 m/r) \right) \; .
\label{eq:minphisol}
\ee
The solution depends on three parameters, $m, a$ and $\alpha$ (see equation
(\ref{eq:parameters})).
When $m = 0$, it represents an isotropic and homogeneous FRW solution first
obtained in Ref. \cite{kn:ellis}.
The static solution ($a=0$) was obtained in Refs. \cite{kn:buchdall,kn:janis}.
The case $\alpha^{2} = 3/4$ (no potential term) was recently analyzed in
\cite{kn:husain}.
\\ \abz
Let us summarize some properties of the obtained solution.
It follows from the proof (see equation (\ref{eq:conf})) that the metric
is conformally static.
As $r \rightarrow \infty$, it asymptotically approaches a spatially flat
Robertson--Walker metric.
If $\alpha = 0$, which corresponds to a massless scalar field in a universe
with a cosmological constant $\Lambda = 3 a^{2}$, the metric is
asymptotically de Sitter.
For any $\alpha$ (except in the case of purely de Sitter universe,
$\alpha = m = 0$) there is a "big bang" singularity as $a T \rightarrow
-\infty$.
If $m \neq 0$, there are also time--like singularities at $r = 0$ and $r =2 m$.
Thus, the solution may be interpreted as an inhomogeneous cosmological model.

\subsection{Conformal coupling}
\label{sec-theorem2}
\abz We now consider the theory described by the action
\be
{S_{c}}[g_{\alpha\beta},\phi] = \int d^{4}\,x\: (-g)^{1/2} \left[
-(16 \pi G)^{-1} R + \frac{1}{2} \phi_{, \alpha} \phi_{, \beta}
g^{\alpha\beta} + \frac{1}{12} R \phi^{2} - {W}(\phi) \right] \; ,
\label{eq:confaction}
\ee
where ${W}(\phi)$ is a potential.
In the absence of the potential term or with a quartic potential, the action
for the scalar field is known to be conformally invariant (see, for instance,
Ref. \cite{kn:birdav}).
The equations obtained by variations of the action (\ref{eq:confaction}) are
more complicated than the ones for a minimally coupled scalar field.
But owing to the theorems proved by Bekenstein \cite{kn:bekenstein1}, one
can generate solutions for the conformal scalar field from known solutions
for the minimal scalar field. The Bekenstein technique, originally formulated
for massless fields, can easily be extended to include potential terms
\cite{kn:abreu}. \\ \abz
Let us summarize here the result: if $(\overline{g}_{\alpha\beta}, \varphi)$
forms a solution of the Einstein--ordinary scalar field equations with a
potential ${V}(\varphi)$ then the Einstein--conformal scalar field equations
with the potential
\be
{W}(\phi) = \varepsilon (1-\zeta^{2} \phi^{2})^{2}
{V}(\frac{\zeta^{-1}}{2} \ln |\frac{1+\zeta \phi}{1-\zeta \phi}|) \label{eq:W}
\ee
are satisfied by the following two sets:
\newcounter{AB}
\begin{list}{\Alph{AB}.}{\usecounter{AB} \setlength{\rightmargin}{\leftmargin}}
\item $\varepsilon = 1$,
\be
\begin{array}{l}
\phi = \zeta^{-1} \tanh (\zeta \varphi) \; , \\
g_{\mu\nu} = \cosh^{2} (\zeta \varphi) \bar{g}_{\mu\nu} \; ; \label{eq:A}
\end{array}
\ee
\item $\varepsilon = -1$,
\be
\begin{array}{l}
\phi = \zeta^{-1} \coth (\zeta \varphi) \; , \\
g_{\mu\nu} = \sinh^{2} (\zeta \varphi) \bar{g}_{\mu\nu} \; . \label{eq:B}
\end{array}
\ee
\end{list}
Here $\zeta = (4 \pi G/3)^{1/2}$. \\ \abz
Applying these mappings to our solution for the ordinary scalar field,
equations (\ref{eq:minmetric}) and (\ref{eq:minphi}), we obtain two solutions
of the Einstein--conformal scalar field equations with the potential
(compare equations (\ref{eq:exppot}), (\ref{eq:parameters}) and (\ref{eq:W}))
\be
{W}(\phi) = \varepsilon (8 \pi G)^{-1} a^{2} (3-4 \alpha^{2})
 |1-\zeta \phi|^{2+2 \sqrt{3} \alpha} |1+\zeta \phi|^{2-2 \sqrt{3} \alpha}
\; . \label{eq:Wmy}
\ee
Thus, we have proved the theorem:
\begin{guess}
If $ds_{o}^{2} = e^{2 u} dt^{2} - e^{-2 u} h_{ij}
dx^{i} dx^{j}$ - is a static vacuum solution of Einstein's equations, then
\be
ds^{2} = \frac{1}{4} (F + \varepsilon F^{-1})^{2} \left\{
\exp (2 \delta u + 8 \alpha^{2} a T) dT^{2} -
\exp (-2 \delta u +2 a T) h_{ij} dx^{i} dx^{j} \right\} \; ,
\label{eq:confmetric}
\ee
\be
\phi = \zeta^{-1} (F^{2} - \varepsilon) (F^{2} + \varepsilon)^{-1} \; ,
\label{eq:conforphi}
\ee
where
\be
F = \exp \left[ (4 \alpha^{2}+1)^{-1/2} u + 2 \alpha a T) 3^{-1/2} \right]
\; , \label{eq:F}
\ee
are a pair ($\varepsilon = \pm 1$) of solutions of the Einstein--conformal
scalar field equations with the potential (\ref{eq:Wmy}), the parameters
$\alpha$ and $\delta$ being related by equation (\ref{eq:parameters}).
\end{guess}
\vspace*{\bigskipamount}
\abz
It is evident that the metrics (\ref{eq:confmetric}) are also conformally
static. When $a = 0$, we get Bekenstein's result \cite{kn:bekenstein1}.
\\ \abz
Let us now consider the case of spherical symmetry. The spherically symmetric
solutions of the Einstein--conformal scalar field equations
are given by expressions (\ref{eq:confmetric})--(\ref{eq:F}),
(\ref{eq:uschw}) and (\ref{eq:hschw}). As in the case of the ordinary scalar
field, each solution (for $\varepsilon = \pm 1$) depends on three parameters,
$m, a$ and $\alpha$. There is also a "big bang" singularity as $a T \rightarrow
-\infty$. When $m = 0, \varepsilon = 1$, our solution reduces to the class of
cosmological solutions found in Ref. \cite{kn:abreu}. \\ \abz
If $m \neq 0$, each solution is singular at both $r = 0$ and $r =2 m$, except
in the case $\alpha = \frac{1}{2 \sqrt{3}}$. The latter case corresponds
(for $a = 0$) to Bekenstein's black hole solution \cite{kn:bekenstein1,
kn:bekenstein2}. In the next section we discuss this case. We will see that,
for $a \neq 0$, $m \neq 0$, the solution describes the formation of a black
hole in a cosmological setting.

\section{Collapsing scalar field}
\label{sec-bh}
\subsection{The solution}
\label{sec-solution}
\abz In this section we shall consider the solution found in the preceding
section corresponding to
$\alpha=\frac{1}{2 \sqrt{3}}$. It is a solution of the Einstein--conformal
scalar field equations with the potential
\be
{W}(\phi) = (H/2 \zeta)^{2} (1+\zeta \phi)^{3} (1-\zeta \phi)
\; , \label{eq:Wbh}
\ee
where we denoted $H=4 a/3$.
The solution written in the ($T,r$)--coordinates reads
\begin{eqnarray}
ds^{2} & = & \frac{1}{4} \left( \varepsilon e^{HT/2} (1-2 m/r)^{1/2}+1
\right)^{2} \nonumber
\\ & & \mbox{} \times \left\{
dT^{2} - e^{HT} \left( \frac{dr^{2}}{(1-2 m/r)} +
 r^{2} d\theta^{2} + r^{2} \sin^{2} \theta
d\varphi^{2} \right) \right\} \; , \label{eq:dsbh}
\end{eqnarray}
\be
\phi = (3/4 \pi G)^{1/2} \left( \varepsilon e^{HT/2} (1-2m/r)^{1/2}-1 \right)
\left( \varepsilon e^{HT/2} (1-2m/r)^{1/2} +1 \right)^{-1} \; ,
\label{eq:phibh}
\ee

To analyze the properties of the solution we first compute the curvature
invariants.
The Ricci scalar is found to be
\be
R_{\alpha}^{\alpha} = - 12 H^{2} \frac{3 \varepsilon e^{HT/2} (1-2m/r)^{1/2}
+1}
{(\varepsilon e^{HT/2} (1-2m/r)^{1/2} +1)^{3}} \; . \label{eq:ricbh}
\ee
If $\varepsilon=+1$, it is bounded for all $r$ and $T$.
If $\varepsilon = -1$, the Ricci scalar blows up  when
$\sqrt{1-2m/r}=e^{-HT/2}$.
\\ \abz
The Weyl tensor can be shown to belong to type D by Petrov \cite{kn:landau},
with the only invariant being
\be
\lambda = \frac{m}{2 r R^{2}} \; , \label{eq:lambda}
\ee
where
\be
R = \frac{1}{2} e^{HT/2} r (\varepsilon e^{HT/2} (1-2m/r)^{1/2} +1)
\label{eq:R}
\ee
is the proper radius of the 2--sphere labeled by ($T,r$).
\\ \abz
It is seen that $\lambda$ blows up as $r R^{2} \rightarrow 0$. The scalar
$R_{\alpha\beta} R^{\alpha\beta}$ has the same kind of singularity.
All the curvature invariants are finite at $r=2m$, therefore, the apparent
singularity of the metric at $r=2m$ can be removed.
We shall do this by introducing the new (isotropic) radial
coordinate,$\bar{r}$,
by the relation:
\be
r = (1+\frac{m}{2 \bar{r}})^{2} \bar{r} \; , \; 0<\bar{r}< \infty \; .
\label{eq:rbar}
\ee
It is also convenient to introduce the new time variable,
\be
\tau = - \varepsilon (2 H)^{-1} e^{HT/2} \; . \label{eq:tau}
\ee
It is evident that the two solutions, with $\varepsilon=+1$ and
$\varepsilon=-1$,
correspond, in fact, to different coordinate patches of the manifold described
below (we assume that $m>0, H>0$ in what follows).
We will see in Sec.\ ~\ref{sec-bh}~\ref{sec-trapped} that the causal structure
of the space--time depends on the value of the dimensionless parameter $mH$
(see also Table 1).
\\
In the ($\tau,\bar{r}$)--coordinates the metric reads
\begin{eqnarray}
ds^{2} & = & \left( 1+\frac{m}{2\bar{r}}-2H \tau
(1-\frac{m}{2\bar{r}}) \right)^{2} \nonumber \\
& & \mbox{} \times \left\{ \frac{d\tau^{2}}{H^{2}\tau^{2}
(1+\frac{m}{2\bar{r}})^{2}} -
H^{2}\tau^{2}(1+\frac{m}{2\bar{r}})^{2} ( dr^2 + r^{2} d\theta^{2} +
r^{2} \sin^{2} \theta d\varphi^{2} ) \right\} \; . \label{eq:bh}
\end{eqnarray}
The metric is now singular ($R=0$) at the surfaces
\be
\bar{r}_{s} = \frac{m}{2} \, \frac{2H\tau+1}{2H\tau-1} \; . \label{eq:rs}
\ee
It is easy to see that the solution is invariant under the transformation:
\be
\begin{array}{l}
\tau \leftrightarrow - \tau \; , \\
\bar{r} \leftrightarrow \frac{m^{2}}{4\bar{r}} \; . \label{eq:inv}
\end{array}
\ee
We shall describe the geometry of space--like surfaces of constant
$\tau$ for $\tau>0$ (see Fig. 1).
\\
\abz
For $0<\tau<(2H)^{-1}$ the spatial sections are asymptotically flat, as
$\bar{r} \rightarrow 0$ and $\bar{r} \rightarrow \infty$.
The two asymptotic regions are connected by a "throat" whose 2--area reaches
 a minimal value at the surface
\be
\bar{r}_{thr} = \frac{m}{2} \sqrt{\frac{1+2H\tau}{1-2H\tau}} \; .
\label{eq:rthr}
\ee
At time $\tau=(2H)^{-1}$, the "right" asymptotic region
($\bar{r} \rightarrow \infty$) shrinks to a singular point and the throat
becomes
infinitely long, with the area decreasing to zero when approaching the "end" of
the throat.
As time develops further, the singularity cuts off more and more of the throat
approaching the "point" $\bar{r}=m/2$ as $\tau \rightarrow \infty$.
As we shall show, an observer in the "left" universe ($0<\bar{r}<m/2$) will
see the appearance of a naked singularity if $0<mH<1/2$.
If $mH>1/2$, the singularity is hidden within an event horizon which appears
to be redshifted to observers outside the horizon.
The horizon "freezes" at $\bar{r}=m/2$ as $\tau \rightarrow \infty$,
but his area increases with the
cosmological expansion of the universe (see equation (\ref{eq:Rb}) bellow).

\begin{figure}[t]
\vspace{-60mm}
   \begin{minipage}[t]{387pt}
  \mbox{\epsfig{file=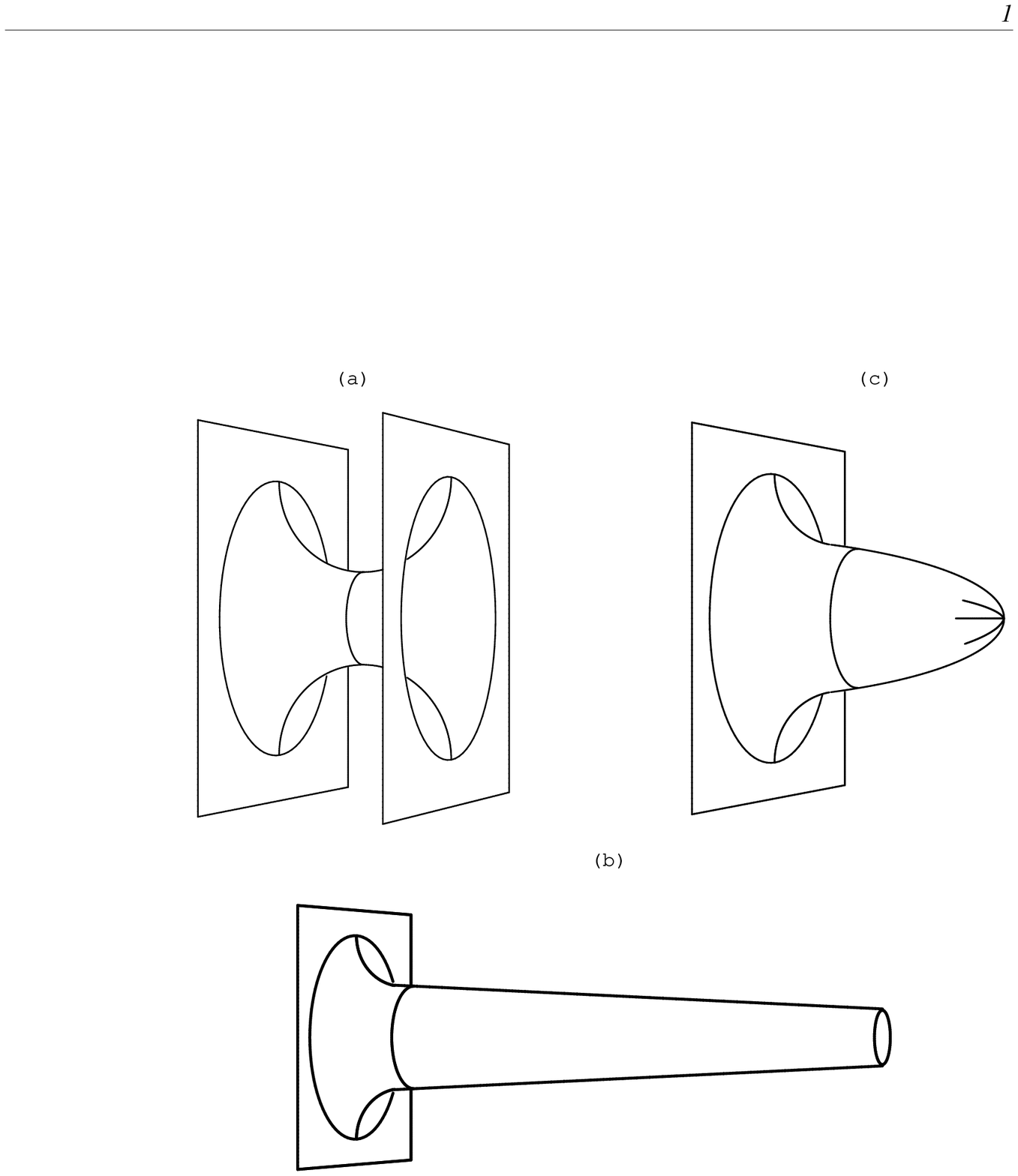,width=387pt}}
\vspace{-50mm}
 \caption{Qualitative representation of the spatial geometry at different
 times $\tau$:
  (a) $0<\tau<(2H)^{-1}$, (b) $\tau = (2H)^{-1}$, (c) $\tau>(2H)^{-1}$.
  One dimension is suppressed. The "bound" photon orbit
  ($\bar{r}=m/2$) is also shown.}
\end{minipage}
\end{figure}

\subsection{Trajectories of test particles}
\label{sec-trajectories}
\abz
Let us consider the motion of test particles in the metric (\ref{eq:bh}).
Since the metric is spherically symmetric, there is one constant of motion
regarded as the angular momentum (per unit rest mass) of a particle,
\be
L = R^{2} \frac{d\varphi}{ds} \; . \label{eq:L}
\ee
Here $s$ is an affine parameter along the particle trajectory, $R$ is the
proper radius (\ref{eq:R}).
In the ($\tau,\bar{r}$)--coordinates,
\be
R^{2} = H^{2} \tau^{2} \bar{r}^{2} (1+\frac{2m}{\bar{r}})^{2} \left(
1+\frac{2m}{\bar{r}}-2H\tau (1-\frac{2m}{\bar{r}}) \right)^{2} \; .
\label{eq:Rinbar}
\ee
Without loss of generality we may restrict attention to study of the equatorial
motion, $\theta=\pi/2$.
Since the metric is conformally static, there is another constant of motion
for massless particles, the "energy" of a photon,
\be
E=\frac{R^{2}}{H^{2} \tau^{2} \bar{r}^{2} (1+\frac{2m}{\bar{r}})^{4}}
\frac{d\tau}{ds} \; .
\label{eq:E}
\ee
Therefore, for isotropic geodesics ($ds=0$) one gets:
\be
\frac{R^{2}}{\bar{r}} \frac{d\bar{r}}{ds} = \pm \sqrt{E^{2} \bar{r}^{2}
(1+\frac{2m}{\bar{r}})^{4} - L^{2}} \; , \label{eq:dr0}
\ee
where the upper sign ($+$) corresponds to photons moving from the "left"
universe towards the "right" universe, the lower sign ($-$) to photons
moving in the opposite direction.
It is easy to show that photons with $L^{2} \leq 4 m^{2} E^{2}$ can pass
through
the throat and travel from one universe to another till the "point" where
they hit the singularity
(equation (\ref{eq:rs}) for $\tau>0$).
The orbit of a photon with $L^{2} \geq 4 m^{2} E^{2}$ will have a "turning"
point, $\bar{r}_{o}$, which obeys the equation:
\be
\bar{r}_{o} (1+\frac{2m}{\bar{r}_{o}})^{2} = |L/E| \; , \label{eq:ro}
\ee
the "light bending" effect.
\\ \abz
The orbits can be explicitly found in two cases: $L=0$ (radial light rays)
and $L^{2} = 4 m^{2} E^{2}$. For radial light rays one easily gets:
\be
\bar{r}_{\ast} - r_{o} = \mp \frac{1}{H^{2} \tau} \; , \label{eq:radial}
\ee
where $r_{o}$ is a constant and
\be
\bar{r}_{\ast} = \bar{r} + m \ln(2\bar{r}/m) - \frac{m^{2}}{4\bar{r}} \; .
\label{eq:rstar}
\ee
If $L=2mE$, there is one unstable "bound" orbit,
\be
\begin{array}{l}
\bar{r} = m/2 \; , \\
\varphi - \varphi_{o} = - (2mH\tau)^{-1} \; , \label{eq:bound}
\end{array}
\ee
with $\tau$ being linear in the affine parameter,
\be
\tau = E s+\tau_{o} \; . \label{eq:taub}
\ee
Note that the orbit (\ref{eq:bound}) is not closed since the proper radius,
(\ref{eq:Rinbar}), is changing with time,
\be
R_{b}^{2} = 4 m^{2} H^{2} \tau^{2} \; . \label{eq:Rb}
\ee
As $\tau \rightarrow 0$, the photon on the orbit (\ref{eq:bound}), which
lies inside the throat, reaches a singularity, $R_{\alpha\beta} R^{\alpha\beta}
\rightarrow \infty$.
Photons with $L=2mE$ moving towards the throat will asymptotically approach
the orbit (\ref{eq:bound}) as $\tau \rightarrow 0$.
The throat between the two universes is, thus, singular at time $\tau = 0$.
\\
\abz
It follows from equations (\ref{eq:E}) and (\ref{eq:dr0}) that, for any $L$ and
$E$, photons moving away from the throat (in both directions) will
asymptotically approach the radial trajectories, (\ref{eq:radial}), as
$\tau \rightarrow 0$. All of them reach (one of) the cosmological horizons,
$R_{DS} = H^{-1}$, within a finite proper time, $s \sim (\tau-\tau_{o})/E$.
We shall discuss the cosmological horizons in Sec.\
{}~\ref{sec-bh}~\ref{sec-coshor}.
\\ \abz
Timelike geodesics cannot be found analytically. It can be shown, however, that
trajectories of massive test particles also approach the asymptotics
(\ref{eq:radial}),
(\ref{eq:rstar}) as $\tau \rightarrow 0$.

\subsection{Trapped surfaces and the black hole event horizon}
\label{sec-trapped}
\abz The causal structure of the space--time reflects its physical properties.
Apparent horizons are surfaces where light wave--fronts are momentarily
"frozen" \cite{kn:hawking}. They are given by the equation:
\be
g^{\alpha\beta} R_{, \alpha} R_{, \beta} = 0 \; , \label{eq:apphorequ}
\ee
where $R$ is the proper radius (\ref{eq:Rinbar}). Equation (\ref{eq:apphorequ})
can be solved exactly in our case. One can get, in the
($\tau,\bar{r}$)--coordinates,
\be
g^{\alpha\beta} R_{, \alpha} R_{, \beta} = \frac{(a_{2} H^{2} \tau^{2} + a_{1u}
H
\tau + a_{0})(a_{2} H^{2} \tau^{2}+ a_{1v} H \tau - a_{0})}{2H^{6} \bar{r}^{6}
(1+\frac{m}{2\bar{r}})^{2} (1+\frac{m}{2\bar{r}}-2H\tau
(1-\frac{m}{2\bar{r}}))^{2}} \; , \label{eq:apphor}
\ee
where
\be
\begin{array}{l}
a_{0} = H^{3} \bar{r} (m^{2} - 4 \bar{r}^{2}) \; , \\
a_{2} = H^{4} (m-2\bar{r})(m+2\bar{r})^{3} \; , \\
a_{1u} = 8 H^{3} m \bar{r}^{2} (2Hm+1) + H^{3} (m-2\bar{r})^{2} (\frac{1}{4}
m^{2} H+2\bar{r}+3Hm\bar{r}+H\bar{r}^{2}) \; , \\
a_{1v} = 8 H^{3} m \bar{r}^{2} (2Hm-1) + H^{3} (m-2\bar{r})^{2} (\frac{1}{4}
m^{2} H-2\bar{r}+3Hm\bar{r}+H\bar{r}^{2}) \; . \label{eq:as}
\end{array}
\ee
The numerator on the right hand side of equation (\ref{eq:apphor}) has in
general
4 roots:
\be
H {\tau_{u\pm}}(\bar{r}) = (-a_{1u} \pm \sqrt{a_{1u}^{2}-4a_{0}
a_{2}})/(2a_{2})
 \; , \label{eq:tauu}
\ee
\be
H {\tau_{v\pm}}(\bar{r}) = (-a_{1v} \pm \sqrt{a_{1v}^{2}+4a_{0}
a_{2}})/(2a_{2})
 \; . \label{eq:tauv}
\ee
We note here that the functions $H \tau_{u\pm}$ and $H \tau_{v\pm}$ are in fact
functions of $H\bar{r}$ and $Hm$. They also possess the symmetry
(\ref{eq:inv}),
 so we again consider the half plane, $\tau>0$.
\\
\abz
The meaning of the functions (\ref{eq:tauu}) and (\ref{eq:tauv}) is that
${\tau_{u\pm}}(\bar{r})$ give the outer apparent horizons and
${\tau_{v\pm}}(\bar{r})$ the inner apparent horizons \cite{kn:hawking}.
We refer to radial light rays whose trajectories obey the equation (compare
equations (\ref{eq:E}) and (\ref{eq:dr0}))
\be
(1+\frac{m}{2\bar{r}})^{2} \frac{d\bar{r}}{d\tau} = \mp \frac{1}{H^{2}
\tau^{2}}
 \; , \label{eq:inoutrays}
\ee
as to outgoing ($-$) and ingoing ($+$) rays, respectively.
\\ \abz
The existence, in general, of two outer and two inner apparent horizons
reflects
the fact that our solution describes the formation of a black hole in a
cosmological setting.
Two (outer and inner) apparent horizons cover the black hole singularity given
by
equation (\ref{eq:rs}) for $\tau>0$.
The other ones surround the "inner" ($0<\bar{r}<m/2$) or "outer" ($\bar{r}>m2$)
cosmological horizons which correspond to $\tau=0$.
We shall say more about the cosmological horizons in the next subsection.
The situation described above is "normal", in the sense that one can clearly
separate out the two phenomena, the cosmological expansion and the collapse
of the scalar field with the formation of a black hole.
If one changes the parameters ($m$ and/or $H$) of the solution, the picture is
also changed.
We summarize here some properties of the apparent horizons (see Table 1).

\begin{table}[t]
\begin{tabular}{||l|l|l|l||}     \hline
parameter      & black hole  & black hole inner  & black hole outer \\
range          & singularity & apparent horizon  & apparent horizon  \\
\hline\hline
$0 < mH < 1/2$ & naked       &  does not         & does not          \\
               &             &  globally exist   & globally exist  \\
\cline{1-3}
$1/2 \leq mH < h_{o}$
               & hidden within &  timelike at      &             \\
\cline{1-1} \cline{4-4}
$h_{o} \leq mH < 3/2$
               & the global  &  late times     & inside the       \\
\cline{1-1} \cline{3-3}
$mH \geq 3/2$  & event horizon & inside the      & event horizon  \\
               &               & event horizon   &                \\ \hline
\end{tabular}
\caption{Some properties of the cosmological black hole solution for different
values of the parameter $mH$}
\end{table}

\abz
The behaviour of the horizons depends on the value of the dimensionless
parameter $mH$.
The functions ${\tau_{v\pm}}(\bar{r})$ are real for all $\bar{r}>0$.
If $mH>1/2$, ${\tau_{v-}}(\bar{r})$ grows from zero (at $\bar{r}=0$), reaches
a maximum at some point $\bar{r}_{m}$, $0<\bar{r}_{m}<m/2$, and then again
decreases to zero (at $\bar{r}=m/2$). The function ${\tau_{v+}}(\bar{r})$
monotonically decreases from infinity (at $\bar{r}=m/2$) to $(4H)^{-1}$
(as $\bar{r} \rightarrow \infty$).
\\ \abz
If $0<mH<1/2$, ${\tau_{v-}}(\bar{r})$ monotonically increases from zero
(at $\bar{r}=0$) to infinity (at $\bar{r}=m/2$), while ${\tau_{v+}}(\bar{r})$
changes from zero (at $\bar{r}=m/2$) to $(4H)^{-1}$
($\bar{r} \rightarrow \infty$).
\\ \abz
If $mH=1/2$, the function ${\tau_{v-}}(\bar{r})$ is negative
(${\tau_{v+}}(\bar{r})$ is positive) for all $\bar{r}>0$, but instead another
inner apparent horizon appears which is situated at $\bar{r}=m/2$. The function
${\tau_{v+}}(\bar{r})$ monotonically grows from zero (at $\bar{r}=0$), reaches
the maximal value $(2H)^{-1}$ at $\bar{r}=m/2$ and then monotonically decreases
to $(4H)^{-1}$ ($\bar{r} \rightarrow \infty$).
\\ \abz
The value $1/2$ of the parameter $mH$ is thus critical for the behaviour of
ingoing light rays. If $mH>1/2$, all ingoing light rays emitted in the vicinity
of the "inner" cosmological horizon first converge and then begin to diverge
due to the cosmological expansion.
The light rays that have passed the throat (at $\bar{r}=m/2$) will be again
converging after some time and they will hit the singularity (where the area of
the light--fronts vanishes) within a finite proper time (affine parameter). \\
\abz
There is another interesting property of the black hole inner apparent horizon
($\tau_{v+}$). It can be shown that it is spacelike everywhere
(for $\bar{r}>m/2$) if $mH>3/2$.
If $1/2<mH<3/2$, there exist a "point", $\bar{r}_{c}$, such that the black hole
inner apparent horizon is timelike for $m/2 \leq \bar{r}<\bar{r}_{c}$.
This means that some ingoing light rays begin to converge before they enter
the black hole region (see the discussion bellow).
\\ \abz
If $0<mH\leq 1/2$, there are some (one if $mH=1/2$) ingoing light--fronts
whose area is non--increasing all the time. Such a behaviour is normal in
asymptotically flat space--times, but it is anomalous in an expanding universe.
\\ \abz
The fact that $mH=1/2$ is a critical value where the solution qualitatively
changes, becomes more evident if one studies the character of the singularity
(see equation (\ref{eq:rs})).
It is easy to show that, if $mH>1/2$, the singularity is spacelike and hidden
within the black hole event horizon. If $mH \leq 1/2$, the singularity becomes
isotropic at some time $\tau_{i}$ ($\tau_{i} \rightarrow \infty$ if $mH=1/2$),
and it is timelike for $\tau>\tau_{i}$. Thus, some observers who have succeeded
to escape to infinity will see the appearance of a naked singularity, if
$0<mH<1/2$.
\\ \abz
To locate the black hole event horizon, we shall consider outgoing radial light
rays. They propagate along the trajectories (see equations (\ref{eq:radial})
and
(\ref{eq:rstar})):
\be
\bar{r} + m \ln(\frac{2\bar{r}}{m}) - \frac{m^{2}}{4\bar{r}} =
\frac{1}{H^{2}\tau}+u \; , \label{eq:outrays}
\ee
where $u$ is a parameter.
The left hand side of equation (\ref{eq:outrays}) is negative for
$0<\bar{r}<m/2$
 and positive for $\bar{r}>m/2$. It is clear that all outgoing light rays with
$u>0$ hit the singularity within a finite time (see equation (\ref{eq:rs}) for
$\tau>0$).
On the other hand, if $mH \geq 1/2$, all outgoing light rays with $u<0$
escape to future null infinity ($\tau \rightarrow \infty$).
Therefore, the {\it black hole event horizon} is given by equation
(\ref{eq:outrays}) with $u=0$.
If $mH \geq 1/2$, the black hole horizon is global and it separates the black
hole region from the rest of the universe: any timelike observer or null ray
in the black hole region inevitably falls to the singularity, while observers
and null rays outside the black hole region can escape to infinity and will
never see the singularity.
If $0<mH<1/2$, there is no global black hole event horizon. For, as we
discussed, some observers at infinity can see the singularity in this case.
Note here that future null infinity ($I^{+}$)
is spacelike since the metric (\ref{eq:bh})
asymptotically ($\bar{r} \rightarrow 0$) approaches the Robertson--Walker one
(see Ref. \cite{kn:hawking}).
The Penrose diagrams for different values of the parameter $mH$ are shown
in Fig. 2.

\begin{figure}[p]
\vspace{-75mm}
   \begin{minipage}[t]{387pt}
  \mbox{\epsfig{file=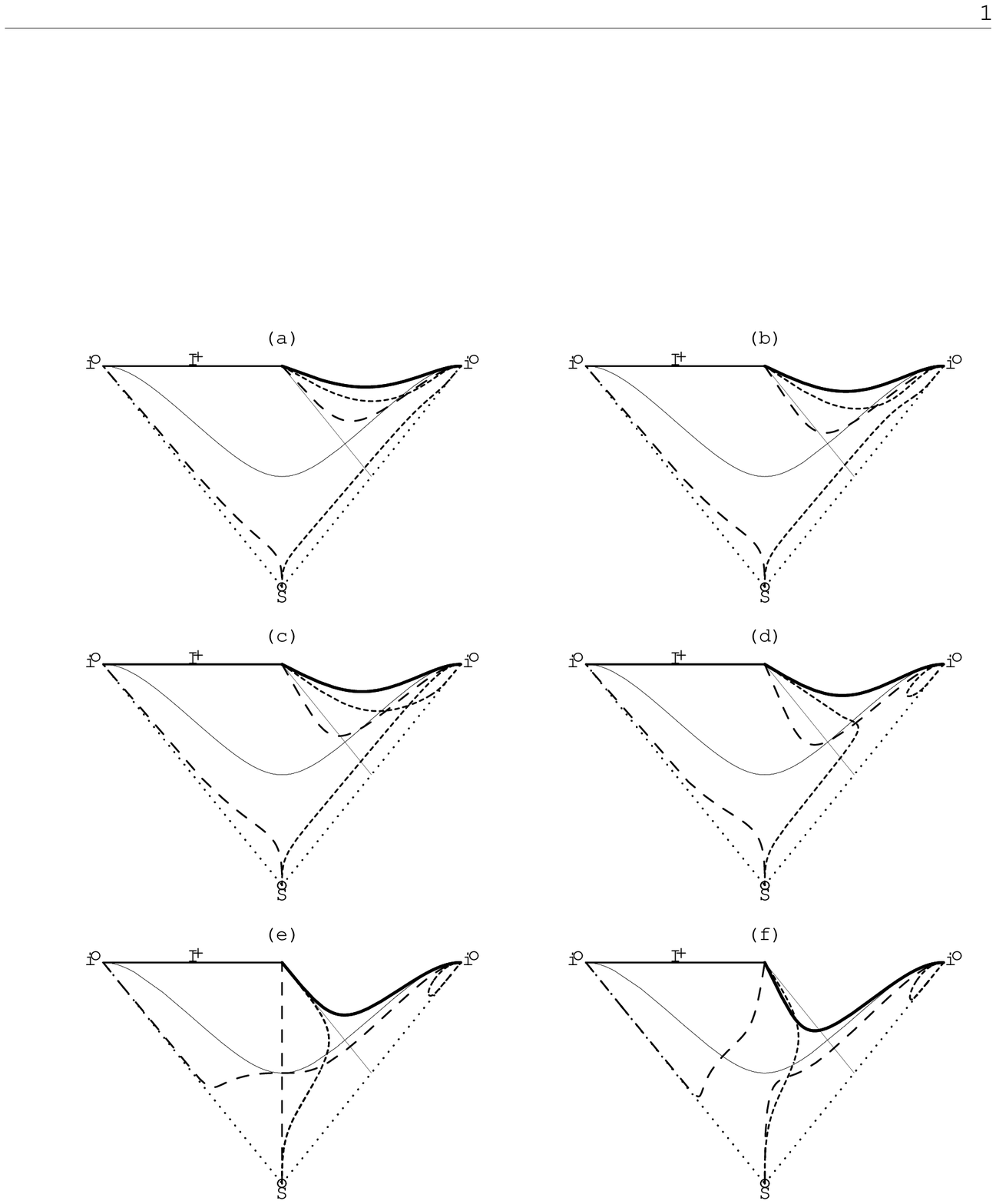,width=387pt}}
\vspace{-50mm}
 \caption{Penrose diagrams of the cosmological black hole solution for
different
values of the parameter $mH$: (a) $mH > 3/2$, (b) $h_{o}<mH<3/2$,
(c) $mH=h_{o}$, (d) $1/2<mH<h_{o}$, (e) $mH=1/2$, (f) $0<mH<1/2$.
  The dashed curves with long segments represent the inner apparent horizons
($\tau_{v\pm}$), the dashed curves with short segments the outer apparent
horizons ($\tau_{u\pm}$). The bold curve is the black hole future singularity,
the point $S$ is the initial ($\tau=0, 0<\bar{r}<\infty$) singularity.
The dotted lines represent the (singular) cosmological horizons.
The solid curve going from the left edge (\i$^o$) to the right one
represents the Cauchy hypersurface $\tau=(2H)^{-1}$. The left and right edges
represent "left" ($\bar{r} \rightarrow 0$) and "right"
($\bar{r} \rightarrow \infty$) spatial infinities, respectively.
The solid line is the black hole event horizon.}
\end{minipage}
\end{figure}

\abz
We shall now consider the outer apparent horizons, equation (\ref{eq:tauu}).
The analysis of the functions ${\tau_{u\pm}}(\bar{r})$ shows that they are
real and positive for all $\bar{r} > m/2$ if $mH \geq h_{o}$, where $h_{o}
\simeq 1.1945$. The function ${\tau_{u-}}(\bar{r})$ grows from zero (at
$\bar{r} = m/2$), reaches its maximal value at some point and then again
decreases to zero ($\bar{r} \rightarrow \infty$), while ${\tau_{u+}}(\bar{r})$
changes from infinity (at $\bar{r} = m/2$) to $(4H)^{-1}$ ($\bar{r}
\rightarrow \infty$). Thus, all outgoing light--fronts emitted in the vicinity
of the "outer" cosmological horizon first decrease and then increase in area.
The light rays with $u>0$ will then again converge and reach the singularity.
Note that the black hole outer apparent horizon ($\tau_{u+}$) is spacelike
and lies inside the event horizon, if $mH \geq h_{o}$. \\ \abz
If $0<mH<h_{o}$, the functions ${\tau_{u\pm}}(\bar{r})$ are real and positive
for $m/2 \leq \bar{r} \leq \bar{r}_{1}$ and for $\bar{r} \geq \bar{r}_{2}$,
where
$H\bar{r}_{1}$ and $H\bar{r}_{2}$ depend on the value of the parameter $mH$.
At $\bar{r}=\bar{r}_{1}$ and $\bar{r}=\bar{r}_{2}$ the function
${\tau_{u-}}(\bar{r})$ smoothly joines ${\tau_{u+}}(\bar{r})$.
Thus, in this case the singularity is not entirely covered by the outer
apparent horizon and there are some (one if $mH=h_{o}$) outgoing light--fronts
whose area is non--increasing all the time.

\subsection{The cosmological horizons}
\label{sec-coshor}
\abz We saw in Sec. ~\ref{sec-bh}~\ref{sec-trajectories} that all ingoing
(outgoing) light rays
reach the inner (outer) cosmolo\-gical horizon, $R_{DS} = H^{-1}$, as $\tau
\rightarrow 0$. Since the metric (\ref{eq:bh}) becomes asymptotically ($\bar{r}
\rightarrow 0$ or $\infty$) de Sitter as $\tau \rightarrow 0$, it would seem
that one could analytically extend the metric beyond the cosmological horizons
($\tau = 0$). As we shall show, this is not the case. \\ \abz
Let us write the metric in the "hyperbolic" coordinates \cite{kn:brill},
($\eta,\chi$). They are given by the relations:
\be
\begin{array}{rll}
\bar{r}_{\ast} & = & \frac{1}{2} \frac{\sin \chi}{\cos(\frac{\eta+\chi}{2})
\cos(\frac{\eta - \chi}{2})} \; , \\
(H^{2} \tau)^{-1} & = & \frac{1}{2} \frac{\sin \eta}{\cos(\frac{\eta+\chi}{2})
\cos(\frac{\eta - \chi}{2})} \; , \label{eq:hyp}
\end{array}
\ee
where $\bar{r}_{\ast}$ is given by equation (\ref{eq:rstar}).
In the ($\eta,\chi$)--coordinates the metric takes the form
\be
ds^{2} = \frac{1}{H^{2} \sin^{2} \eta} (d\eta^{2} - d\chi^{2}) -
R^{2} (d\theta^{2} + \sin^{2} \theta d\varphi^{2} ) \; , \label{eq:dshyp}
\ee
where $R$ is now a function of $\eta$ and $\chi$,
\be
{R^{2}}(\eta,\chi) = \frac{\sin^{2} \chi}{H^{2} \sin^{2} \eta}
\frac{\bar{r}^{2} (1+\frac{m}{2\bar{r}})^{4}}{\bar{r}^{2}_{\ast}} \; .
\label{eq:Rhyp}
\ee
The cosmological horizons correspond to $\eta \pm \chi = \pi (2 n +1)$,
$n=0,\pm1,\cdots$.
Unfortunately, the function ${R}(\eta,\chi)$ is not analytic across the
horizons. Indeed, from equations (\ref{eq:hyp}),(\ref{eq:Rhyp}) one can get,
 for example ($\chi \neq \pi n$):
\be
\frac{1}{R} \frac{\partial R}{\partial \chi} \rightarrow \mp \frac{m}{2}
\cot \chi \ln (\cos(\frac{\eta \pm \chi}{2})) \; ,
\mbox{when $\eta \pm \chi \rightarrow \pi (2 n+1)$.} \label{eq:dRchi}
\ee
Therefore, some components of the Riemann tensor (which involve the derivatives
of ${R}(\eta,\chi)$) blow up, though all the curvature invariants remain
finite at the cosmological horizons (see the discussion after equation
(\ref{eq:R})). We shall prove here that such a situation occurs in any
coordinate system, namely, some components of the Riemann tensor along
geodesics blow up when the geodesics approach one of the cosmological
horizons (this is a kind of the "parallelly propagated curvature singularity"
\cite{kn:wald} or null singularity). In fact, we shall show that the metric
is only $C^{0}$ across the cosmological horizon.
Owing to the spherical symmetry it is
sufficient to consider the following form of the metric:
\be
ds^{2} = {g_{ab}}(x^{c}) dx^{a} dx^{b} - {R^{2}}(x^{c}) (d\theta^{2} +
\sin^{2} \theta d\varphi^{2} ) \; , a,b,c=0,1 \; . \label{eq:dsgen}
\ee
Consider any timelike or null geodesic, ${x^{\alpha}}(s)$, and suppose that
the 2--metric $g_{ab}$ is analytic across the horizon. Thus, $k^{\alpha}
\equiv \frac{dx^{\alpha}}{ds}$ is finite there. For simplicity we assume
the motion is equatorial ($\theta=\pi/2$). One can prove then that
the following relation holds:
\be
k^{\alpha} k^{\beta} R_{\alpha\varphi\beta\varphi} = R \frac{d^{2}R}{ds^{2}} \;
{}.
 \label{eq:2dprop}
\ee
Equation (\ref{eq:2dprop}) can be proved by considering, for example, the
geodesic deviation equation \cite{kn:landau}. \\ \abz
On the other hand, it was mentioned in Sec.
{}~\ref{sec-bh}~\ref{sec-trajectories} that
all particles approach the horizon along the trajectories (\ref{eq:radial}),
(\ref{eq:rstar}), with $\tau$ being linear in the affine parameter,
$\tau \sim E (s-s_{o})$. For such trajectories one gets (see equation
(\ref{eq:Rinbar})):
\be
R \sim H^{-1} \mp mH\tau \ln|mH^{2}\tau| + {O}(\tau^{2} \ln \tau) \; .
\label{eq:Rasymp}
\ee
{}From equations (\ref{eq:2dprop}) and (\ref{eq:Rasymp}) we then obtain:
\be
k^{\alpha} k^{\beta} R_{\alpha\varphi\beta\varphi} \sim \mp \frac{mHE}{s-s_{o}}
+
 {O}(\ln|s-s_{o}|) \; . \label{eq:Rsing}
\ee
Therefore, the components $R_{a\varphi b\varphi}$ (and $R_{a\theta b\theta}$)
blow up at the cosmological horizon. An extended body will rotate with an
infinite angular velocity when it approaches the cosmological horizon.

\section{Concluding remarks}
\label{sec-conclusion}
\abz
One of the results presented in this work is the cosmological black hole
solution which is discussed in Section ~\ref{sec-bh}. It displays some
interesting features. At earlier times ($0<\tau<(2H)^{-1}$) there are two
universes connected by a "throat" which is traversable to light rays and
observers. One of the universes then collapses and a future singularity is
formed (for $\tau < 0$ the solution describes a white hole in the cosmological
setting). Depending on the value of the parameter, $mH$, there are two
possibilities: either some observers in the second (expanding) universe will
see the naked singularity ($0<mH<1/2$) or the singularity will be hidden
within a global event horizon ($mH \geq 1/2$) which appears to be redshifted
to external observers.
\\
\abz
In the first case cosmic censorship \cite{kn:penrose} is manifestly violated.
It is not a serious violation yet since the model we have considered is not
generic. For instance, the stress--energy tensor obtained from the action
(\ref{eq:confaction}) does not satisfy the (strong) energy condition
\cite{kn:hawking}. Note, however, that another counterexample to the cosmic
censorship conjecture was recently obtained in Ref. \cite{kn:kastor}.
\\ \abz
Another interesting feature of our solution is that the metric is not smooth
across the cosmological horizons (if $m \neq 0$). The singularity is rather
mild: radial timelike and null geodesics can pass through it without being
affected, but an extended body will be torn to pieces by fast rotation.
The lack of smoothness at some horizons was also observed in other exact
solutions
(see Refs. \cite{kn:kastor, kn:chrusciel}).
In our case, this can be understood along the following line of arguments.
In the theory
described by the action (\ref{eq:confaction}), one can define the effective
gravitational constant,
\be
G_{eff} = G (1-\frac{4\pi}{3} G \phi^{2})^{-1} \; . \label{eq:Geff}
\ee
It is easy to show that $G_{eff} \rightarrow \infty$ as $\tau \rightarrow 0$
(compare equations (\ref{eq:phibh}), (\ref{eq:rbar}), (\ref{eq:tau}) and
(\ref{eq:Geff})). On the other hand, we saw that the throat is "visible" to
distant observers as a massive body (the "light bending" effect, etc.).
As the throat becomes singular at $\tau = 0$, the observers will feel its
"tail" via the Newtonian potential.
If the "mass parameter" $m=0$, the (single) cosmological horizon is regular.
In the latter case, the solution describes a homogeneous isotropic universe
with a "big bang" (or "big crash") singularity.

\section*{Acknowledgments}
\abz
I am grateful to Professor Jacob Bekenstein for suggesting me the problem,
for stimulating discussions during the course of this work, for careful
reading an early draft of the manuscript and for many helpful comments.
Valuable discussions with Dr. Renaud Parentani are also appreciated.
The exact solutions given in this paper were verified and the pictures
were produced using the symbolic computation package MATHEMATICA. I would
like to express my thanks to Professor Gerald Horwitz for having introduced
me to the world of MATHEMATICA and for allowing me to use his computer.
I acknowledge partial support derived from a grant of the Israel Science
Foundation, administered by the Israel Academy of Sciences.


\begin{thebibliography}{99}
\bibitem{kn:hawking} S. W. Hawking and G. F. R. Ellis, {\em The Large Scale
Structure of Space--Time}, (Cambridge University Press, Cambridge, 1973).
\bibitem{kn:tolman} R. G. Tolman, Proc. Nat. Acad. Sci. US {\bf 20}, 169
(1934).
\bibitem{kn:vaidya} P. C. Vaidya, Proc. Indian Acad. Sci. A{\bf 33}, 264
(1951);
Nature {\bf 171}, 260 (1953).
\bibitem{kn:choptuik} M. Choptuik, Phys. Rev. Lett. {\bf 70}, 9 (1993).
\bibitem{kn:abraham} A. Abraham and C. R. Evans, Phys. Rev. Lett. {\bf 70},
2980 (1993).
\bibitem{kn:halliwell} J. J. Halliwell, Phys. Lett. B{\bf 185}, 341 (1987).
\bibitem{kn:barrow} J. D. Barrow, Nucl. Phys. B{\bf 296}, 697 (1988); \\
J. D. Barrow and S. Gotsakis, Phys. Lett. B{\bf 214}, 515 (1988);
B{\bf 258}, 299 (1988); \\
S. Gotsakis and P. J. Saich, Class. Quantum Grav. {\bf 11}, 383 (1994); \\
A. B. Burd and J. D. Barrow, Nucl. Phys. B{\bf 308}, 929 (1988).
\bibitem{kn:feinstein} A. Feinstein and J. Ib\'{a}\~{n}ez, Class. Quantum Grav.
{\bf 10}, 93 (1993); ibid, L227 (1993); \\
J. M. Aguirregabiria, A. Feinstein, and J. Ib\'{a}\~{n}ez, Phys. Rev. D{\bf
48},
4662, 4669 (1993).
\bibitem{kn:bekenstein1} J. D. Bekenstein, Ann. Phys. {\bf 82}, 535 (1974).
\bibitem{kn:abreu} J. P. Abreu, P. Crawford, and J. P. Mimoso, Class. Quantum
Grav. {\bf 11}, 1919 (1994).
\bibitem{kn:landau} L. D. Landau and E. M. Lifshitz, {\em The Classical Theory
of Fields} (Pergamon Press, Oxford, 1971).
\bibitem{kn:buchdall} H. Buchdahl, Phys. Rev. {\bf 115}, 1325 (1959).
\bibitem{kn:janis} A. I. Janis, D. C. Robinson, and J. Winicour, Phys. Rev.
{\bf 186}, 1729 (1969).
\bibitem{kn:yano} K. Yano, {\em Integral Formulas in Riemannian Geometry}
(Pure and Applied Mathematics: A Series of
Monographs and Textbooks) (Marcel Dekker, Inc., New York, 1970).
\bibitem{kn:ellis} G. F. R. Ellis and M. S. Madsen, Class. Quantum Grav.
{\bf 8}, 667 (1991).
\bibitem{kn:husain} V. Husain, E. A. Martinez, and D. N\'{u}\~{n}ez,
{\em Preprint} gr--qc/9402021.
\bibitem{kn:birdav} N. D. Birrell and P. C. W. Davies, {\em Quantum Fields
in Curved Space} (Cambridge University Press, Cambridge, 1982), p. 44.
\bibitem{kn:bekenstein2} J. D. Bekenstein, Ann. Phys. {\bf 91}, 75 (1975).
\bibitem{kn:brill} D. R. Brill and S. A. Hayward, Class. Quantum Grav. {\bf
11},
359 (1994).
\bibitem{kn:wald} R. M. Wald, {\em General Relativity} (The University of
Chicago Press, Chicago and London, 1984), p. 215.
\bibitem{kn:penrose} R. Penrose, in {\em General Relativity, an Einstein
Centenary Survey}, ed. S. W. Hawking and W. Israel (Cambridge University
Press, Cambridge, 1979).
\bibitem{kn:kastor} D. R. Brill, G. T. Horowitz, D. Kastor and J. Traschen,
Phys. Rev. D{\bf 49}, 840 (1994).
\bibitem{kn:chrusciel} P. T. Chru\'{s}ciel and D. B. Singleton, Commun.
Math. Phys. {\bf 147}, 137, (1992).
\end{thebibliography}
\end{document}